\documentclass{aa}
\usepackage{graphicx}
\usepackage{amsmath,epsfig,rotating,amssymb}

\begin{document}

\def \rot{{\rm {\bf rot} }}
\def \grad{{\rm {\bf grad} }}
\def \div{{\rm div}}
\def \cha{\widehat}
\def \pr{{\it permanent}  regime }
%\textwidth 6.in
%\oddsidemargin 0.in
%\evensidemargin 0.in
%\topmargin 0.in
%\headheight 0.5in
%\headsep 0.2in
%\textheight 9.in
%\parindent 0.2in
%\pagestyle{myheadings}

%\input{psfig.tex}

\author{Hennebelle P. \inst{1}, Teyssier R. \inst{2}}

\institute{Laboratoire de radioastronomie millim{\'e}trique, UMR 8112 du
CNRS, 
\newline {\'E}cole normale sup{\'e}rieure et Observatoire de Paris,
 24 rue Lhomond, 75231 Paris cedex 05, France 
\and Service d'Astrophysique, CEA/DSM/DAPNIA/SAp, Centres d'\'Etudes de Saclay, 
91191 Gif-sur-Yvette Cedex, France }

\offprints{  P. Hennebelle \\
{\it patrick.hennebelle@ens.fr}   }

\title{Magnetic processes in a collapsing dense core. II-Fragmentation}

\subtitle{Is there a fragmentation crisis ? }

\abstract
{A large fraction of stars are found in binary systems. It is therefore important for our understanding 
of the star formation process, to investigate the fragmentation of  dense molecular cores.}
{We study the influence of the magnetic field, ideally coupled to the gas,
 on the fragmentation in multiple systems  of collapsing cores.}
{We present high resolution numerical simulations performed with the RAMSES MHD code starting 
with a uniform sphere in solid body rotation and a uniform magnetic field parallel to the rotation 
axis. We pay particular attention to the strength of the magnetic field and interpret the results
using the analysis presented in a companion paper.  }
{The results depend much on the amplitude, $A$, of the perturbations seeded initially.
For a low amplitude, $A=0.1$, we find that for values of the mass-to-flux over critical mass-to-flux ratio, $\mu$, as high as $\mu = 20$, 
the centrifugally supported disk which fragments in the hydrodynamical case, is stabilized 
and remains axisymmetric. 
Detailed investigations reveals that this is due to the rapid growth of the 
toroidal magnetic field induced by the differential motions within the disk. 
For  values of $\mu$ smaller $\simeq 5$, corresponding 
to larger magnetic intensities, there is no centrifugally supported disk because of 
magnetic braking.
When the amplitude of the perturbation is equal to $A=0.5$, each initial peak develops independently
and the core  fragments for a large range of $\mu$. Only for values of $\mu$ close 
to $1$ is the magnetic field able to prevent the fragmentation. }
{Since a large fraction of stars are binaries, the results of low magnetic intensities preventing 
the fragmentation in case of weak perturbations, is problematic. We discuss three possible mechanisms
which could lead to the formation of binary systems namely the presence of large amplitude fluctuations in the 
core initially, the ambipolar diffusion and the fragmentation during the second collapse. }

\maketitle

\section{Introduction}

Understanding the fragmentation of  collapsing prestellar dense cores, is of great importance 
to our understanding of the star formation process. In particular, determining the 
number of fragments, their  masses and their orbital characteristics are fundamental 
and rather challenging problems. Most of stars are 
found in binary or multiple systems (Duquenoy \& Mayor 1991).
Many studies have investigated this issue in the context of hydrodynamical calculations
(e.g. Miyama 1992, Boss 1993, Bonnell 1994, Truelove et al. 1998, Bate \& Burkert 1997, 
Bodenheimer et al. 2000, Matsumoto \& Hanawa 2003, Hennebelle et al. 2004, Goodwin et al. 2004,
Banerjee et al. 2004). 
Although detailed conclusions appear to be very sensitive to the 
initial   conditions (i.e. core shape, thermal, rotational and turbulent energy  
as well as to the  equations of state), some trends can nevertheless be inferred. 
It is widely accepted that, under realistic initial conditions, a collapsing dense core 
fragments into a few objects, the exact number depending on the specific conditions. Therefore, 
fragmentation of a rotating collapsing dense core appears to be the most widely accepted 
mechanisms to explain the formation of multiple systems. Indeed, it is today the only 
viable mechanism since other possibilities such as fission of a rotating protostar or
 capture of a companion, fail in realistic conditions to produce a large fraction of 
binaries (Bodenheimer et al. 2000).

Due to the difficulty of the problem and despite its importance, the question of the role 
of the magnetic field in this process has remained little addressed. 
With the recent progress achieved in numerical techniques as well as the increasing 
computing power, several studies have recently investigated this issue.
 Hosking \& Whitworth (2003) using an SPH two fluids code, conclude that magnetically 
subcritical cores do not fragment.  Machida et al. (2005) using a nested grid code performed an 
extensive number of calculations varying the initial core rotation and magnetic field.
They find that fragmentation is possible if the rotation is sufficiently large and the magnetic
field strength sufficiently small.  
Fragmentation of a collapsing magnetized cloud using adaptive mesh refinement techniques have also 
been studied by Ziegler (2005), Banerjee \& Pudritz (2006) and Fromang et al. (2006) which all find that 
the magnetic field has a strong influence. 
Price \& Bate (2007), using  magnetized SPH techniques
which insure the nullity of  the divergence of the magnetic field, find that  
with large initial perturbations,  even for large values of the magnetic strength, the 
fragmentation is possible.
In all these studies, it has been found that the magnetic field has a strong 
impact on the fragmentation of the collapsing dense core.

\setlength{\unitlength}{1cm}
\begin{figure}
\begin{picture}(0,18)
%\put(0,12){\includegraphics[width=7cm]{mu1000/dens_vit_xy00010.ps}}
%\put(0,6){\includegraphics[width=7cm]{mu1000/dens_vit_xy00015.ps}}
%\put(0,0){\includegraphics[width=7cm]{mu1000/dens_vit_xy00027.ps}}
\put(0,0){\includegraphics[width=7cm]{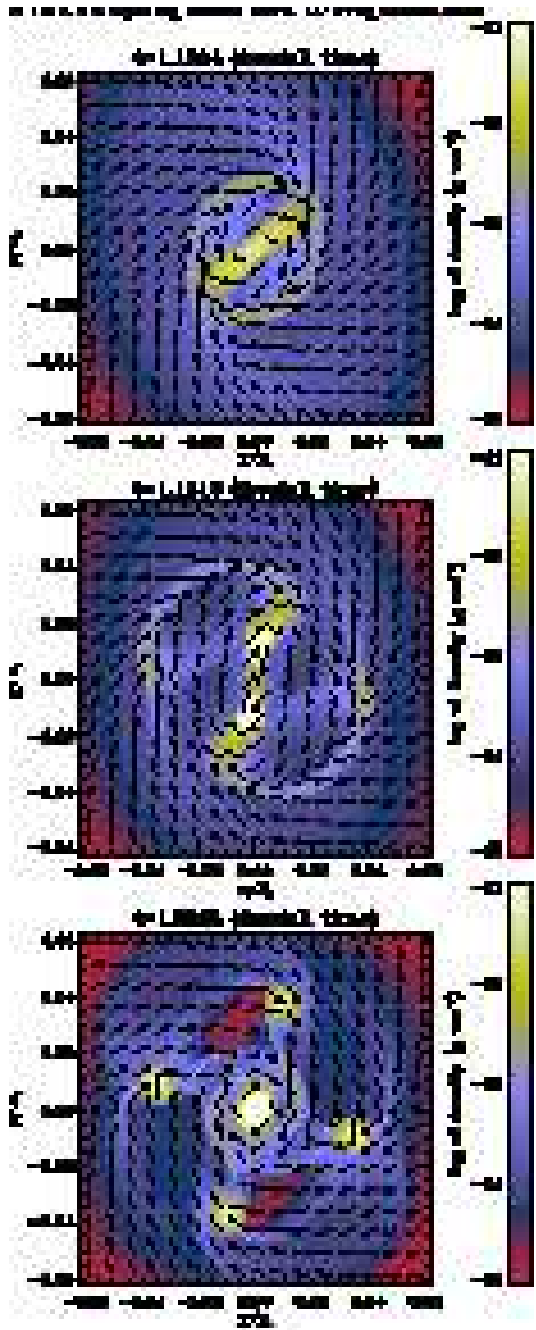}}
\end{picture}
\caption{Equatorial density and velocity field for mass-to-flux over critical 
mass-to-flux ratio $\mu=1000$ and perturbation amplitude equal to $A=0.1$.
The size of the figure corresponds to about one tens of the cloud initial radius.
In physical units, this is about 300 AU. The freefall time is about $3 \times 10^4$ years.}
\label{mu1000_Ap0.1}
\end{figure}

\setlength{\unitlength}{1cm}
\begin{figure}
\begin{picture}(0,18)
%\put(0,12){\includegraphics[width=7cm]{mu50/dens_vit_xy00010.ps}}
%\put(0,6){\includegraphics[width=7cm]{mu50/dens_vit_xy00020.ps}}
%\put(0,0){\includegraphics[width=7cm]{mu50/dens_vit_xy00040.ps}}
\put(0,0){\includegraphics[width=7cm]{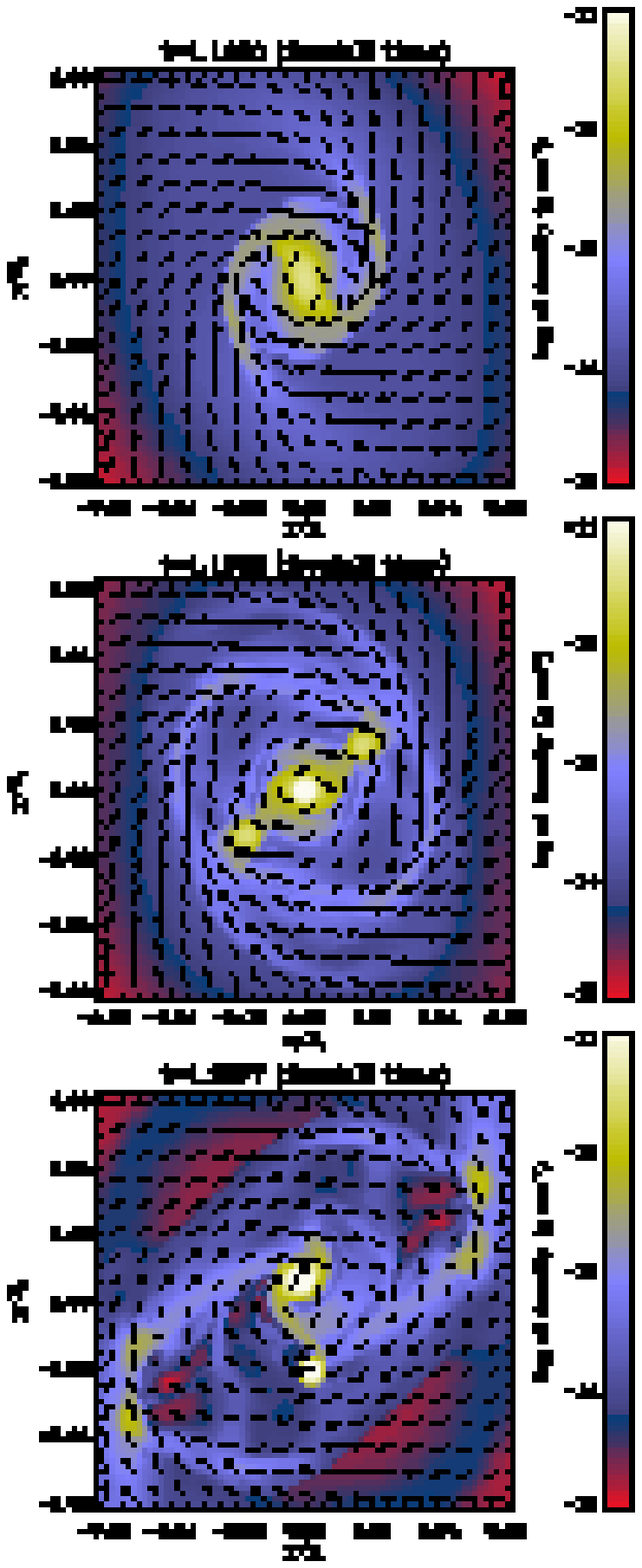}}
\end{picture}
\caption{Same as Fig.~\ref{mu1000_Ap0.1} for $\mu=50$ and $A=0.1$.}
\label{mu50_Ap0.1}
\end{figure}

In this paper, we study further the fragmentation of collapsing cores, focusing mainly 
on the influence of the magnetic field strength and the amplitude of the initial perturbation. 
Using the analysis developed 
in the companion paper by Hennebelle \& Fromang (2007), thereafter paper I, 
we try to  understand, when possible quantitatively, the physical reasons for the
various numerical results.   
As in paper I, we consider initially a uniform density cloud 
in solid body rotation,
 threaded by a uniform magnetic field. At this stage, we restrict the problem to the case 
where magnetic field and rotation axis are aligned. We note that Price \& Bate (2007) have 
investigated the influence of initially perpendicular rotation axis and magnetic field. 
More realistic initial conditions, including non uniform density field, turbulent velocity field 
and rotation axis non aligned with the magnetic field, will be considered in future studies.   
In particular, we determine the lowest 
 value of $\mu$, the  mass-to-flux over critical mass-to-flux ratio for which fragmentation 
is suppressed.  It turns out that the results depend strongly on the 
amplitude of the initial perturbation. 
 
The paper is organized as follow. In the first section, we consider the same initial 
conditions as used in paper I, namely a spherical uniform one solar mass dense core.
The  thermal  over gravitational energy of the core is equal to 
$\simeq$0.37 and the rotation over gravitational energy is 0.045. 
Note that such rotation are typical of  (may be slightly larger than)  values observed in dense cores (Goodman et al. 1993).
The initial density is about $\simeq \, 5   \times 10^{-18}$ g cm$^{-3}$
and the cloud radius, $R_0 \simeq$0.016 pc. The freefall time is thus of the order of
$3 \times 10^4$ years.  
We use a barotropic equation of state:
$C_s^2 = (C_s^0)^2 \times (1 + (\rho/\rho_c)^{4/3})^{1/2}$, 
where   $C_s \simeq 0.2$ km/s is the sound speed and
 $\rho_c=10^{-13}$ g cm$^{-3}$.
This set of cloud parameters is known to
give rise to a disk that unambiguously fragments in the pure hydrodynamics case.  The only difference 
with paper I is that 
 an $m=2$ perturbation of amplitude $A=0.1$ in the density field, 
$\rho(r,\theta,z) = \rho_0 \times (1 + A {\rm cos} (m \theta) )$, 
 as well as in the $B_z$ field, 
is added. In the second section, we further discuss the physical interpretation of the results 
obtained in our simulations and we estimate analytically the critical value of $\mu$ for 
which it is expected that the magnetic field stabilizes the disk. 
In the third section, we consider initial density perturbations of amplitude 0.5 and show that in this 
case, fragmentation can be obtained for a much wider range of $\mu$. 
The fourth section  provides a discussion on possible mechanisms leading to disk fragmentation and the 
formation of binaries, even in the presence of magnetic fields.
The sixth section concludes the paper.

\setlength{\unitlength}{1cm}
\begin{figure}
\begin{picture}(0,18)
%\put(0,12){\includegraphics[width=7cm]{mu20/dens_vit_xy00010.ps}}
%\put(0,6){\includegraphics[width=7cm]{mu20/dens_vit_xy00055.ps}}
%\put(0,0){\includegraphics[width=7cm]{new_pdt_mu20.ps}}
\put(0,0){\includegraphics[width=7cm]{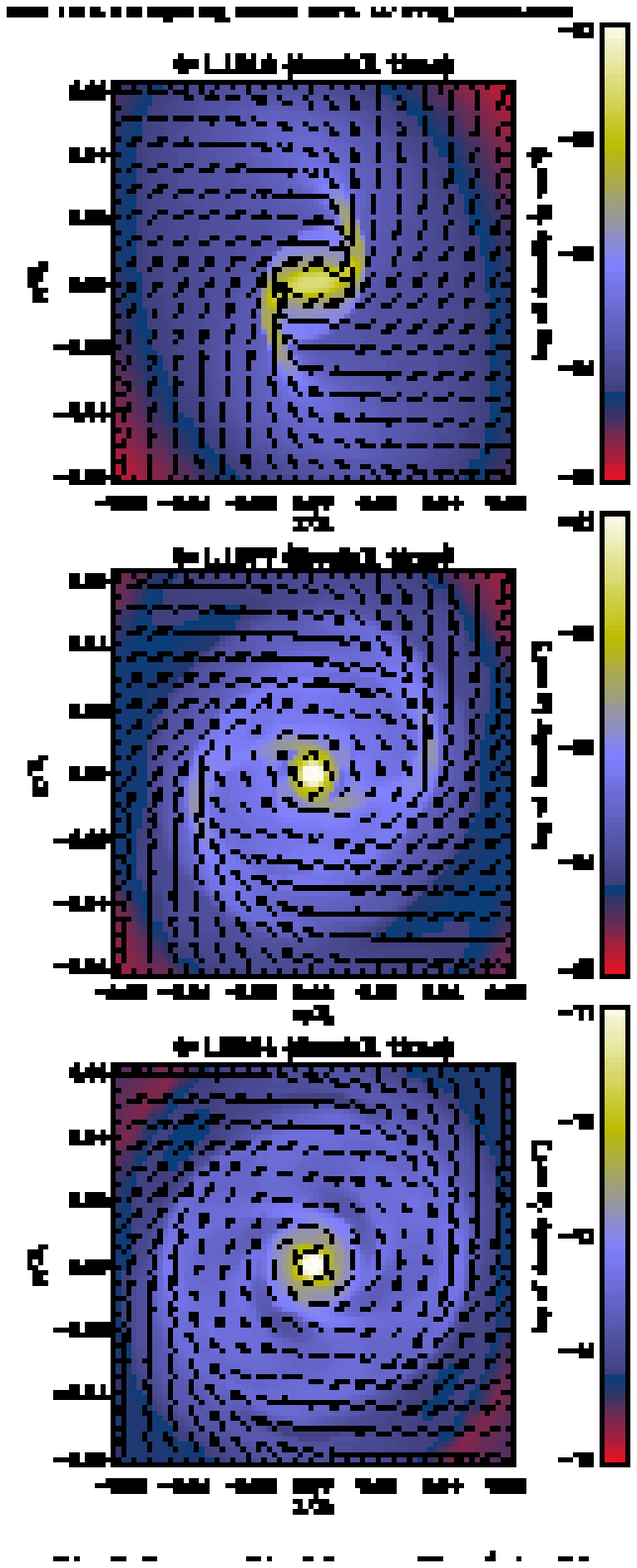}}
\end{picture}
\caption{Same as Fig.~\ref{mu1000_Ap0.1} for $\mu=20$ and $A=0.1$.}
\label{mu20_Ap0.1}
\end{figure}

\setlength{\unitlength}{1cm}
\begin{figure}
\begin{picture}(0,12)
%\put(0,6){\includegraphics[width=7cm]{mu5/dens_vit_xy00020.ps}}
%\put(0,0){\includegraphics[width=7cm]{mu5/dens_vit_xy00082.ps}}
\put(0,0){\includegraphics[width=7cm]{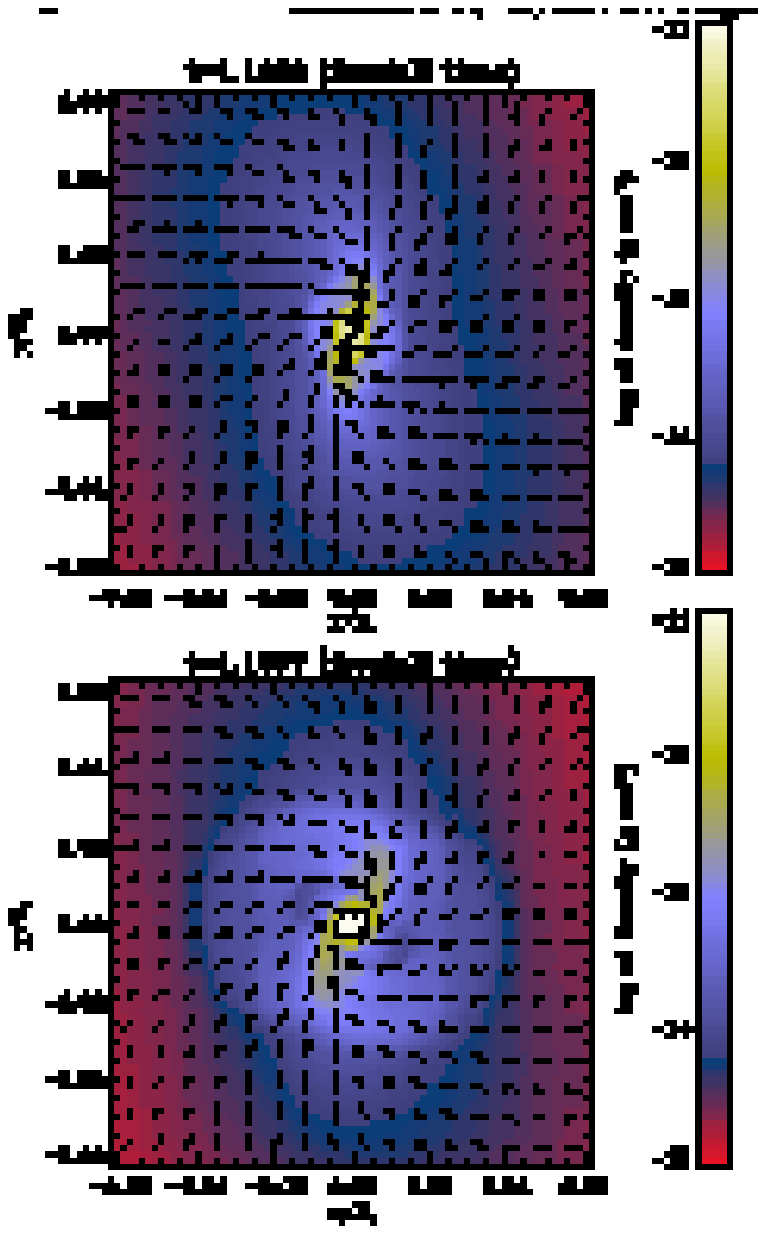}}
\end{picture}
\caption{Same as Fig.~\ref{mu1000_Ap0.1} for $\mu=5$ and $A=0.1$.}
\label{mu5_Ap0.1}
\end{figure}

\section{Weak initial perturbations}
We investigate the fragmentation in the case where the amplitude of the perturbations 
is initially $A=0.1$. In this case it is found that without rotation no fragmentation 
occurs. Therefore, we call this type of initial condition, rotationally driven fragmentation.
We next consider various values of the magnetic mass-to-flux over critical mass-to-flux ratio,
namely $\mu=1000$ (quasi hydrodynamical case), $\mu=50$, 20, 5, 2 and 1.25. Recalling that 
 $\mu=1$ corresponds to the case of a cloud supported by the magnetic field, the 
last value corresponds to a strongly magnetized supercritical cloud. 

In the following, we display  the inner part of the collapsing cloud. The 
size of each plot is about $\simeq R_0/10$. This corresponds to a physical size of 
$1.6 \times 10^{-3}$ pc or about $\simeq$300  AU. 

\subsection{Hydrodynamic  case: fragmentation}
We present the hydrodynamical  case  ($\mu=1000$).
Figure~\ref{mu1000_Ap0.1} shows three snapshots.
Equatorial density and  velocity field are displayed. 
 The first snapshot is taken shortly after the formation 
of the first Larson core. A rotationally supported structure has formed and 
since it is very unstable, a strong spiral pattern is developing. This is very 
similar to the results obtained by other authors (e.g. Matsumoto \& Hanawa 2003, 
Hennebelle et al. 2004, Goodwin et al. 2004). The next snapshot shows that 
the central bar-like structure, fragments into two objects whereas the spiral 
pattern keeps growing and expanding due to further accretion of mass and momentum.
The last snapshot shows that the centrifugally supported structure continues
fragmenting and that 5 fragments have developed.  Altogether, this is very 
similar to the results reported by many authors investigating the fragmentation 
of hydrodynamical rotating cores. We note that in spite of the strongly non-linear 
evolution, the result remains remarkably symmetric due to proper resolution of 
the Jeans length (10 numerical cells per Jeans length).

\subsection{Weak magnetic field cases: suppression of fragmentation}
We present results for the weak field cases, $\mu=50$ and $\mu=20$.
Figure~\ref{mu50_Ap0.1} shows results for $\mu=50$. Due to the difference in the 
initial magnetic strengths, it is not possible to compare  the simulations at 
exactly the same times. 
The first snapshot shows a spiral pattern which is  similar to the 
one seen in the previous 
case. A difference however, is that the central bar-like structure appears to be 
much shorter. This is likely due to the central magnetic field which is strongly 
amplified by the rapid twisting of field lines. As a result, the cloud first fragments
into three fragments instead of two. This is due to the fact that  since the central 
bar-like structure does not fragment in two objects as in the previous case, more material
and angular momentum is available in its vicinity  to produce one object on each side. 
The next snapshots show that a symmetry breaking occurs due to one of the two 
satellites having merged with the central more massive fragment. The structure in the 
outer part ($x>0.03 R_0$) is still quite symmetric. Further fragments are forming in 
the outer part of the spiral pattern at $x=\pm0.05 R_0$. 

From these results, one concludes  that even for values  of $\mu$ as
large  as 50,  the  magnetic field  has  a significant  impact on  the
evolution of the  centrifugally supported inner structure particularly
 its fragmentation.  This is  due to the strong amplification of the
toroidal  and  radial magnetic  field  generated  by the  differential
motions  in the  collapsing core  and in  the  centrifugally supported
structure (see paper I and following sections).

Figure~\ref{mu20_Ap0.1} shows three snapshots for $\mu=20$.
The first snapshot shows a weak spiral pattern in the inner part. 
The second and third panels shows that the centrifugally supported structure grows 
as in the previous cases but it remains much more uniform and the spiral 
pattern is much less pronounced. 
As shown in 
paper I, the angular momentum profile is similar to the hydrodynamical case since the 
 magnetic braking is very weak during the collapse phase.  
Therefore, the  differences with the case $\mu=50$, is not smaller 
angular momentum but rather a stronger magnetic field. 
Indeed, with a magnetic toroidal component,  
the velocity of the fast MHD wave, which can be loosely seen as a 
effective sound speed, is: 
$\sqrt{C_s^2 + V_a^2}$ where $V_a^2 = B_{\theta}^2 / 4 \pi \rho$
(see section~\ref{interp}). Therefore, the disk is stabilized 
against gravitational fragmentation. 
As a result, no fragmentation is obtained, only one central star forms and 
grows by accretion. We stress that the value of the magnetic field 
corresponding to $\mu=20$,  is  very modest,   far below the 
 values inferred from observations which indicates that $\mu=1-2$ (Crutcher 1999).

Finally, to explore whether the suppression of the fragmentation depends on 
initial conditions, such as the values of $\alpha$ and $\beta$ (the ratios 
of thermal and rotational over gravitational energies), we have performed runs 
with different $\alpha$ and $\beta$ values.
First, we explore the influence of stronger and weaker rotation choosing  $\beta=0.2$
and $\beta=0.02$ respectively keeping the value of $\alpha$ constant. 
The behaviour of these runs is very similar to the one with $\beta=0.045$ 
presented here. A large disk forms but it remains well axisymmetric and does not fragment.
We have also explored the effect of a smaller thermal energy, taking $\alpha=0.2$
and keeping $\beta=0.045$. The disk is a little more axisymmetric than in the case 
$\alpha=0.37$ but still does not fragment either. 

We conclude that, for small initial density perturbations, a magnetized dense core
ideally coupled to the magnetic field and having $\mu=20$ does not fragment 
for a large range of initial conditions. 

It is worth to compare with the study of Machida et al. (2005) although it is
 not straightforward because their initial
conditions consist in a filament and therefore are different from ours. 
Also they used different definitions to quantify the 
amount of rotation and magnetic energies in their simulations. Comparison with their Fig.~10 is 
worthwhile nevertheless. Their parameter $\omega$ turns out to be equal to $\sqrt{\beta}$.
However, their initial state has a peak density of about $10^4$ g cm$^{-3}$ which is less
dense than ours by a factor of roughly 100. Since the ratio of rotational over gravitational energies increases during 
the collapse, we can say that our $\beta = 0.045$ correspond to $\omega < 0.2$ in Machida et al.
study (for a homologous contraction $\beta \propto 1/r \propto \rho^{-1/3}$, this would indicate that 
$\beta$ should be divided by roughly $\simeq 4-5$ to be compared with Machida et al.'s results).
For  values of 
$\omega < 0.2 $, they found that for $B_z / \sqrt{8 \pi C_s^2 \rho_c} > 0.1$, they have no fragmentation. 
Comparing this value with our parameters is again not straightforward but 
$B_z / \sqrt{8 \pi C_s^2 \rho_c}$ is about $\simeq 1 / \sqrt{\alpha}  \mu$. For $\alpha = 0.37$ 
and $\mu=20$, $ 1 /  \sqrt{\alpha} \mu  \simeq 0.1$. Therefore, our results broadly agrees with the results 
of Machida et al. (2005).

Comparison with the study of Price \& Bate (2007) is not possible at this stage, since they use 
a barotropic equation of state which becomes adiabatic at $10^{-14}$ g cm$^{-3}$. Thus, even the hydrodynamical 
calculations they present does not fragment when the perturbation is weak (their figure 3). 

\subsection{Intermediate magnetic  intensity: formation of pseudo-disks}

We present results for the intermediate magnetized cases, namely
$\mu=5$ and  $\mu=2$.  Figure~\ref{mu5_Ap0.1} shows  two snapshots for
$\mu=5$. As shown in paper I, no  centrifugally supported structure
forms instead a magnetized pseudo-disk develops. 
Pseudo-disks  arise when disk--like  structures form, say
oblate ellipsoids, which  are not supported by rotation  but rather by
 magnetic support.  According to the analysis presented
in paper  I, the angular momentum  is  lower for these values of
$\mu$  primarily because   the  collapse occurs  first  along the
field lines. Thus, the   material within  the pseudo-disk and  central object
was initially  located along the  pole and has less  angular momentum.
Some  magnetic braking  also  occurs,   reducing the  angular
momentum further.   As  a  consequence,  no centrifugally  supported  disk  is
observed and no fragmentation is occurring.

Figure~\ref{mu2_Ap0.1} shows two snapshots for $\mu=2$.
The first panel shows that a filamentary structure has developed after
$t=1.51$ freefall times. We believe that this filament is due to the non-linear
evolution of the initial $m=2$ perturbation. The reason why it has a shape
different than in the previous case is that since the cloud is more supported by the 
magnetic field, the collapse  lasts 30$\%$ longer. Therefore the perturbation  
has more time to develop and to become non-linear. Indeed, if no perturbation is initially
included, an axisymmetric  pseudo-disk develops.

\subsection{Nearly critical core}
We now consider the case of a very magnetized supercritical core having 
$\mu=1.25$.  As expected the collapsing time is now longer and roughly
equal to two freefall times. 
As revealed by the two panels of  Fig.~\ref{mu1.25_Ap0.1}, the
collapsing dense core remains almost axisymmetric in spite of the 
initial $m=2$ density perturbations. This is due to the strong magnetic support
which prevents the development of the perturbation. Indeed, since the cloud is nearly
supported by the magnetic field, the magnetic Jeans Mass is just slightly smaller 
than the cloud mass. Therefore, each peak of the $m=2$ density perturbation is 
gravitationally stable. Moreover, unlike the thermal support, the magnetic support 
does not decrease during the collapse since the ratio of magnetic over gravitational 
energy stays roughly constant. Therefore the magnetic Jeans mass stays roughly 
constant,  unlike the thermal Jeans mass.

\setlength{\unitlength}{1cm}
\begin{figure}
\begin{picture}(0,12)
%\put(0,6){\includegraphics[width=7cm]{mu2/dens_vit_xy00060.ps}}
%\put(0,0){\includegraphics[width=7cm]{mu2/dens_vit_xy00089.ps}}
\put(0,0){\includegraphics[width=7cm]{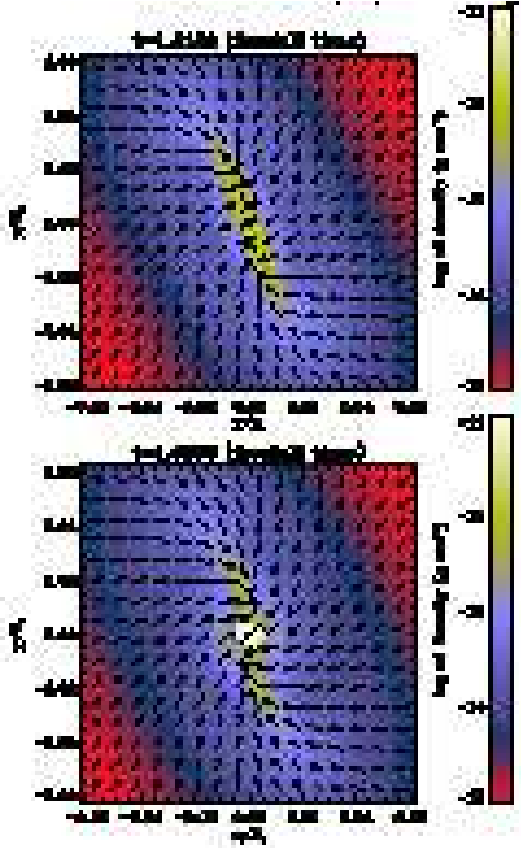}}
\end{picture}
\caption{Same as Fig.~\ref{mu1000_Ap0.1} for $\mu=2$ and $A=0.1$.}
\label{mu2_Ap0.1}
\end{figure}

\setlength{\unitlength}{1cm}
\begin{figure}
\begin{picture}(0,12)
%\put(0,6){\includegraphics[width=7cm]{mu1.25/dens_vit_xy00070.ps}}
%\put(0,0){\includegraphics[width=7cm]{mu1.25/dens_vit_xy00094.ps}}
\put(0,0){\includegraphics[width=7cm]{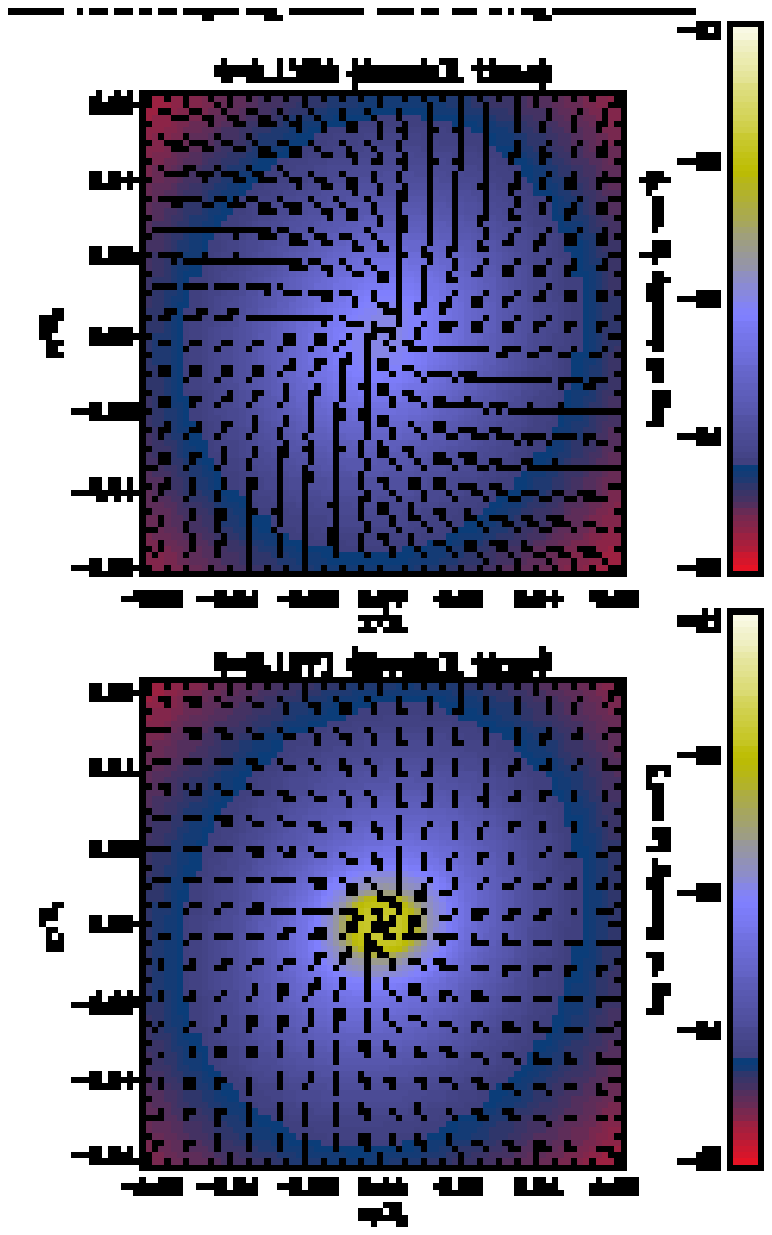}}
\end{picture}
\caption{Same as Fig.~\ref{mu1000_Ap0.1} for $\mu=1.25$ and $A=0.1$.}
\label{mu1.25_Ap0.1}
\end{figure}

\section{Physical interpretation}
\label{interp}

\setlength{\unitlength}{1cm}
\begin{figure}
\begin{picture}(0,12)
\put(0,8){\includegraphics[width=7cm]{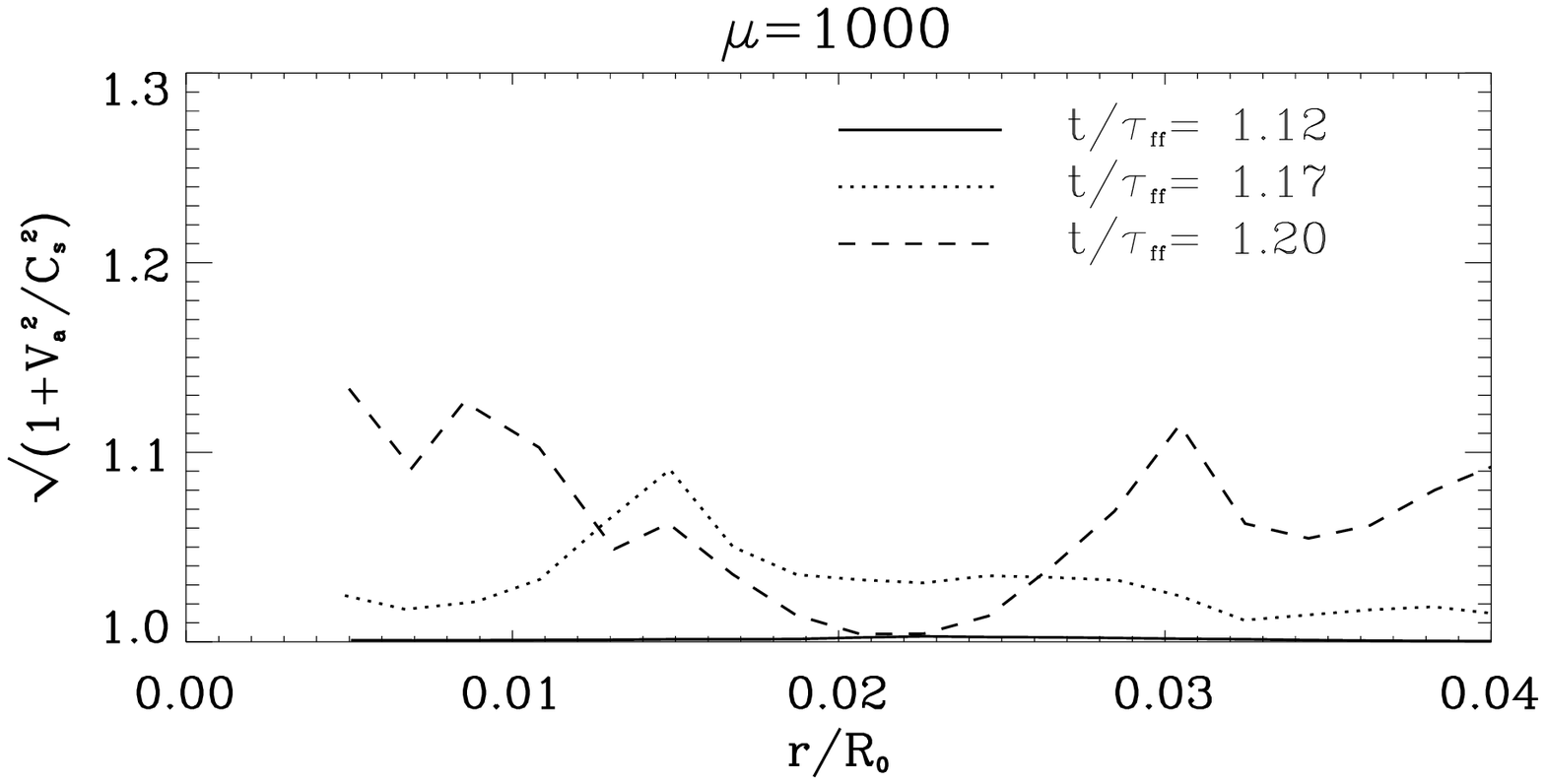}}
\put(0,4){\includegraphics[width=7cm]{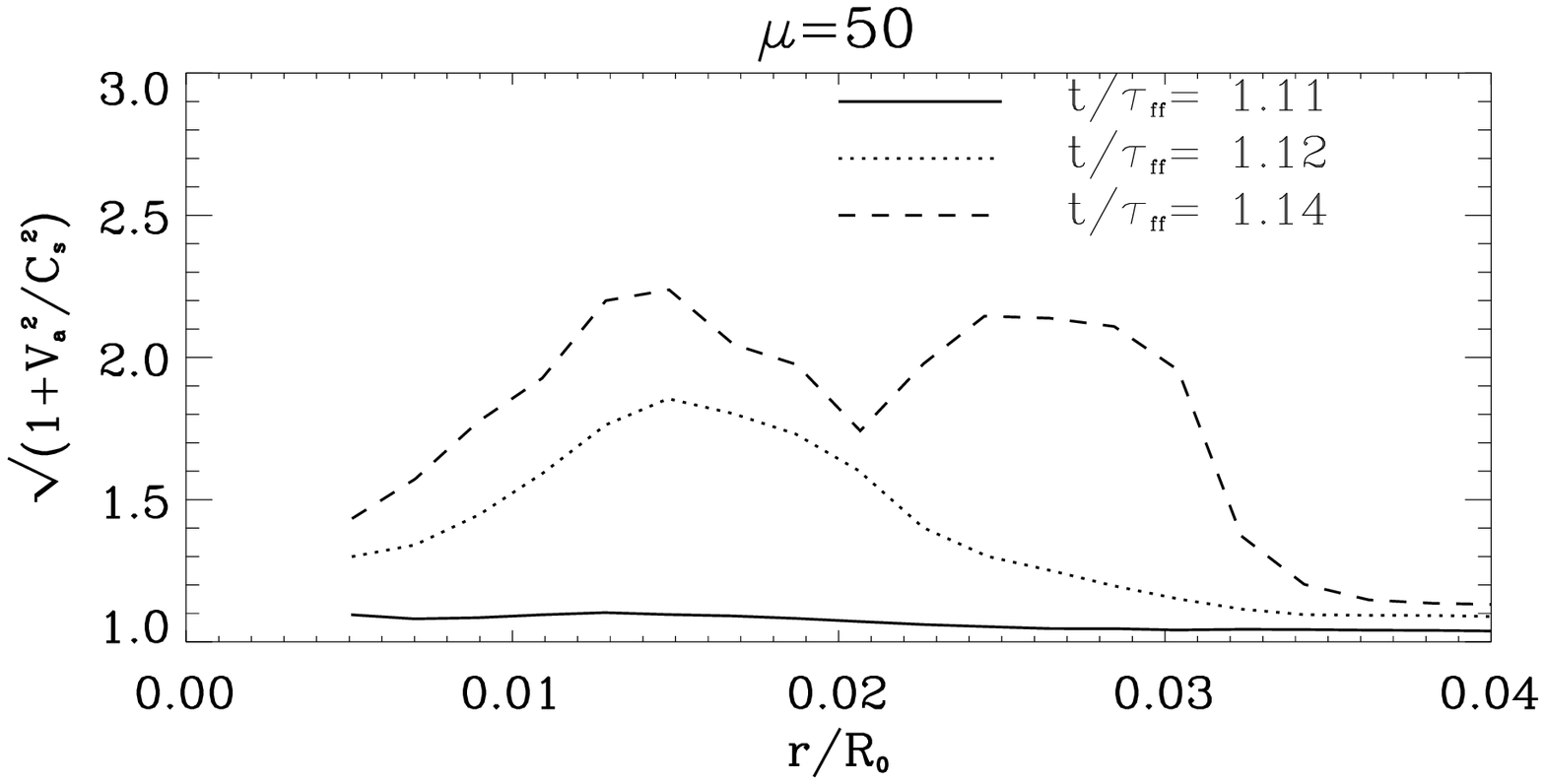}}
\put(0,0){\includegraphics[width=7cm]{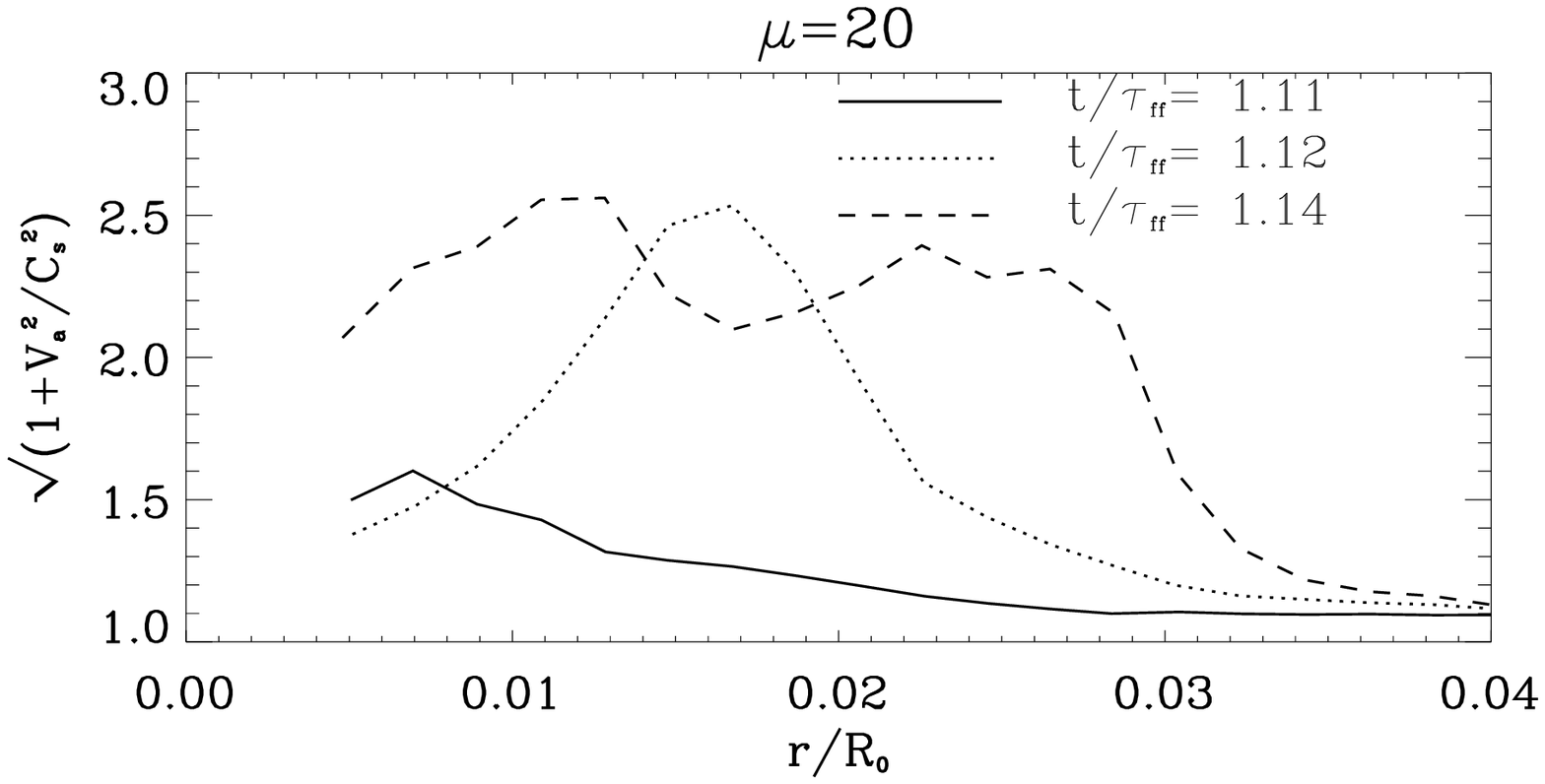}}
\end{picture}
\caption{$\sqrt{1 + V_a^2/C_s^2}$ within the disk.}
\label{va_cs}
\end{figure}

One of the important conclusions reached in the previous section is that
fragmentation can be suppressed even  for large values of $\mu$, i.e. weak
magnetic fields.  
Here we discuss the physical reason for the disk stabilization 
arising at low magnetic strength. We stress that 
since the angular momentum is nearly identical for $\mu=20$ to
 the hydrodynamical case (see paper I), the reason of this 
stabilization is not magnetic braking.

\subsection{Influence of the magnetic field on the disk stability}
The  influence   of  the  magnetic   field  on  the  stability   of  a
self-gravitating  disk,  has first  been  investigated by  Lynden-Bell
(1966) for  a uniform  rotation. The  dispersion relation  he obtained
(see his Eq.~1)  entails an effective sound speed  $\sqrt{C_s^2 +
V_a^2}$  showing that  the magnetic  pressure term   has  a stabilizing
influence.  However, there is also  a  destabilizing contribution due to the magnetic
tension term.  As a result, such configurations are unstable.

Elmegreen (1987) and Gammie (1996) both consider the influence of 
the magnetic field on the stability of a differentially rotating 
system. In that case, the shear drives the growth of a 
toroidal magnetic component which stabilizes the disk. 
They both conclude that whereas in the absence of shear  the magnetic 
field is strongly destabilizing, it has a stabilizing influence when 
significant shear is present.  More precisely, Gammie (1996) computes the response 
 of the disk to nonaxisymmetric perturbations for various value of $Q$, the 
Toomre parameter. He shows that the response is much weaker when a 
substantial magnetic field is present and 
concludes that the stabilizing effect of the field 
becomes significant once the magnetic pressure is comparable to the gas 
pressure. This is qualitatively in good agreement with our results. In particular, we see
that the disk in case $\mu=20$ remains much more axisymmetric than 
in the case $\mu=1000$ and $\mu=50$.

In order to  quantitatively verify that this mechanism is responsible of the disk stabilization, 
we have computed the Alfv\'en speed associated with the azimuthal component of the magnetic field,
$B_\theta/\sqrt{4 \pi \rho}$,  within the disk 
at various times corresponding to the period during which fragmentation is  taking place.
Since both $B_\theta$ and $\rho$ vary, we have integrated  at every radius along the z-axis 
through the disk. 
As explained in paper I, as soon as the azimuthal component of the magnetic field exceeds the
ram pressure of infalling material, a magnetic tower builds up and the disk starts expanding.
In this case, the disk boundaries are not clearly defined.  
We adopt here a density threshold to define the disk and stop the integration when $\rho < 10^{-14}$ g cm$^{-3}$. 
We calculate 
$<V_a^2> = \int B_\theta^2 / 4 \pi \rho \times \rho dz / \int \rho dz $, 
which allows us to estimate $\sqrt{1 + V_a^2/C_s^2}$ within the disk.
Figure~\ref{va_cs} shows $\sqrt{1 + V_a^2/C_s^2}$
for $\mu=1000$, 50 and 20. As can be seen, for the case $\mu=1000$, the quantity
$\sqrt{1 + <V_a^2>/C_s^2}$ remains smaller than 1.2 at all the 
 times displayed.
Figure.~\ref{mu1000_Ap0.1} shows that by this time, fragmentation has already occurred. 
The second and third panel of Fig.~\ref{va_cs} show that  $\sqrt{1 + <V_a^2>/C_s^2}$ grows 
much more rapidly for smaller $\mu$ reaching values greater than 2. The growth is 
 more rapid for $\mu=20$  than it is for $\mu=50$.
 We conclude that the growth of the toroidal 
component induced by the differential rotation, is mainly responsible, 
for   disk stabilization.

For simplicity, the  works mentioned above, have restricted the analysis to a
thin disk. However,  the expansion of the magnetic tower  
triggered by the growth of the toroidal magnetic field, 
 removes some material from the disk and therefore
reduces the disk surface density. This makes the disk even  less prone to fragment. 
This effect which is not taken into account in the thin disk analysis, 
certainly contributes to further stabilize the centrifugally supported structure against
fragmentation. Indeed, the mass of the disk and the mass within the tower, turn out to be roughly 
comparable. 

\subsection{Analytical estimate of the critical value of $\mu$}

We estimate analytically 
the value of $\mu$ at which one expects to find a significant influence 
of the magnetic field  on the fragmentation. Following Gammie (1996), we 
will consider that a strong stabilizing influence is achieved when the 
Alfv\'en speed associated to the toroidal component is comparable to the sound speed. 

The principle of the analysis is as follow: we compute the growth rate
of   the  toroidal  component   inside  the   centrifugally  supported
structure,  so that  we  can estimate  the  time, $\tau_{{\rm  mag}}$,
needed for the Alfv\'en speed to become comparable to the sound speed.
We then  compute the dynamical  time of the disk,  $\tau_{{\rm dyn}}$,
over  which fragmentation  will occur.  The critical  value  of $\mu$,
below which  fragmentation is  quenched, is obtained  when $\tau_{{\rm
dyn}}$ is equal to $\tau_{{\rm mag}}$.

The growth time of the toroidal magnetic component within the disk has two
contributions. First the twisting of the radial component due to  differential 
rotation proportional to $V_\theta B_r / r$ and second the wrapping of the vertical magnetic 
component due to the vertical  gradient of $B_z$ proportional to $V_\theta B_z / z$, 
where $z$ is  the disk height. The  divergence constraint
 shows that $B_r/ r \simeq B_z / z$, indicating that the 2 contributions are comparable.

Thus,  the  growth  of  $B_\theta$  can be  estimated  by  writing
$\partial_t B_\theta  \simeq B_z V_\theta /  h$ where $h$  is the disk
height.   Since we  are  interested  in computing  the  growth of  the
magnetic field  from  initially small values, we  neglect the influence
of the  magnetic tower  on the disk  vertical expansion and  write: $h
\simeq \sqrt{C_s^2 / (4 \pi G \rho)}$. With Eq.~(1) of paper I, we get
$h \simeq r / \sqrt{2 d} $, where $d$ is the ratio of the density over
singular isothermal density.
  
If we assume that the disk is roughly Keplerian, we get 
$V_\theta \simeq \sqrt{M_s G /  r} $, $M_s$ being the mass of 
the star and the disk. 
Taking the expression  of $B_z$ stated by  Eqs.~(1)  of paper I, we obtain: 
\begin{eqnarray}
B_\theta \simeq t \times { \sqrt{2d} H_z \sqrt{M_s} C_s^2  \over  r^{5/2} }
\end{eqnarray}

The value of $\tau_ {{\rm mag}}$, is obtained by requiring
$(B_\theta / \sqrt{4 \pi \rho})  / C_s \simeq 1 $.
Using the expression of density stated by Eq.~(1) of paper I, we obtain:
\begin{eqnarray}
\tau_{{\rm mag}} \simeq { r^{3/2} \over    H_z \sqrt{G M_s}    }. 
\label{growingtime}
\end{eqnarray}

The dynamical time of the disk, i.e. the time relevant for fragmentation, is the 
rotation time:
\begin{eqnarray}
\tau_{{\rm dyn}} \simeq { 2 \pi r^{3/2} \over \sqrt{G M_s} }.
\end{eqnarray}
Therefore, we obtain:
\begin{eqnarray}
{\tau_{{\rm mag}} \over \tau _{{\rm dyn}}} \simeq {1  \over 2 \pi  H_z}.
\end{eqnarray}
To compute the value of $H_z$, one  can use Table~1 of paper I, but it
is more convenient to express the result in terms of the initial cloud
parameters  namely $\mu$  and $\alpha$.   The latter  is the  ratio of
thermal over  gravitational energy  and  is  given by $\alpha  = 5/2
C_s^2 R_0  / (M G) $.   From Eq.~4 of paper  I, the value  of $H_z$ is
given by $\sqrt{G}B_z^0 R_0 /  (2 C_s^2)$, whereas from the definition
of  $\mu$, $B_z^0$  can  be written  as  $M_0 /  (\pi R_0^2)  \sqrt{G}
\sqrt{9  \pi^2   /  5}  /   (0.53  \mu)$.   Gathering   the  different
expressions, we get:
\begin{eqnarray}
{\tau_{{\rm mag}} \over \tau _{{\rm dyn}} }  \simeq {  \sqrt{2}  \over 3 \sqrt{5} \pi  }
   \times  \mu \alpha.
\end{eqnarray}
We can deduce the critical value of $\mu$
\begin{eqnarray}
\mu_c \simeq \frac{15}{\alpha}
\end{eqnarray}
Note that this value is approximate and should be valid within a factor of a few.

For  $\alpha  \simeq 0.37  $,  we  obtain  $\mu_c \simeq  40$.   This
analytical estimate  is therefore in reasonable agreement  with our
numerical  simulation, since for $\mu=20$, fragmentation is  suppressed. 
In this case, the toroidal
magnetic pressure becomes comparable to  the sound speed in about half
a rotation period.  Since  typically fragmentation is occurring over a few
rotation period, such a fast  growth time appears to be sufficient to
prevent  the  disk  from  fragmenting.   
On the other hand for  $\mu=50$  the  disk  is
fragmenting, confirming that  this case is slightly above the  critical value.  

For $\mu=1000$,  one has  $\tau_{{\rm mag}}  /  \tau _{{\rm
dyn}} \simeq 20$.  Therefore in  such a  case, the growth of the toroidal
component is too slow to significantly influence the disk evolution.

\setlength{\unitlength}{1cm}
\begin{figure}
\begin{picture}(0,18)
%\put(0,12){\includegraphics[width=7cm]{mu20_0.5/dens_vit_xy00010.ps}}
%\put(0,6){\includegraphics[width=7cm]{mu20_0.5/dens_vit_xy00020.ps}}
%\put(0,0){\includegraphics[width=7cm]{mu20_0.5/dens_vit_xy00036.ps}}
\put(0,0){\includegraphics[width=7cm]{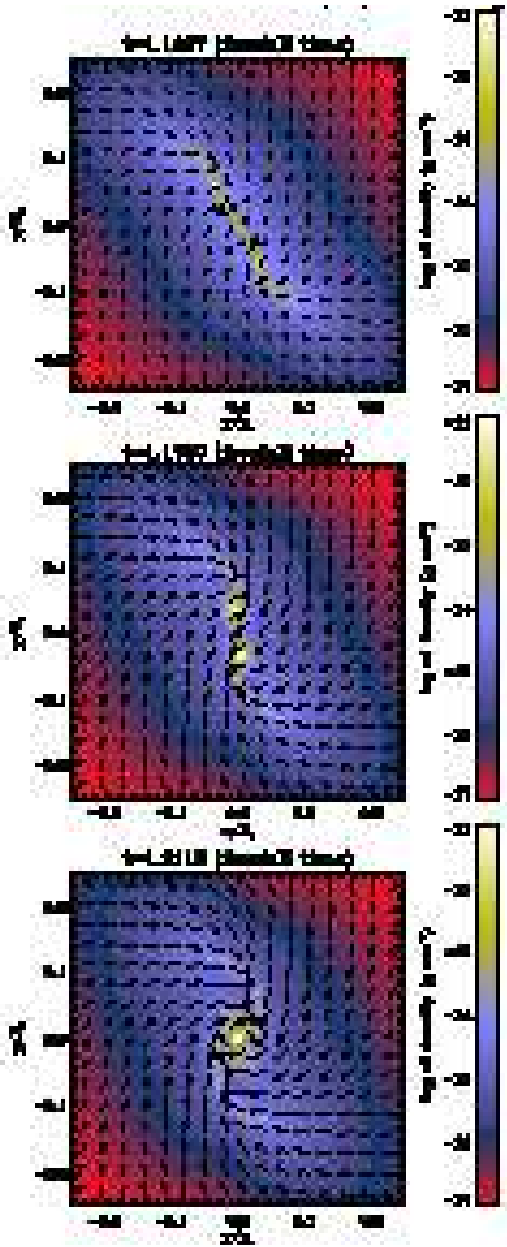}}
\end{picture}
\caption{Same as Fig.~\ref{mu1000_Ap0.1} for $\mu=20$ and $A=0.5$.
Note nevertheless that the figure corresponds to a  length of $\simeq$1200 AU, 
four times bigger than the length  displayed in Fig.~\ref{mu1000_Ap0.1}. }
\label{mu20_Ap0.5}
\end{figure}

\setlength{\unitlength}{1cm}
\begin{figure}
\begin{picture}(0,12)
%\put(0,6){\includegraphics[width=7cm]{mu2_0.5/dens_vit_xy00040.ps}}
%\put(0,0){\includegraphics[width=7cm]{mu2_0.5/dens_vit_xy00072.ps}}
\put(0,0){\includegraphics[width=7cm]{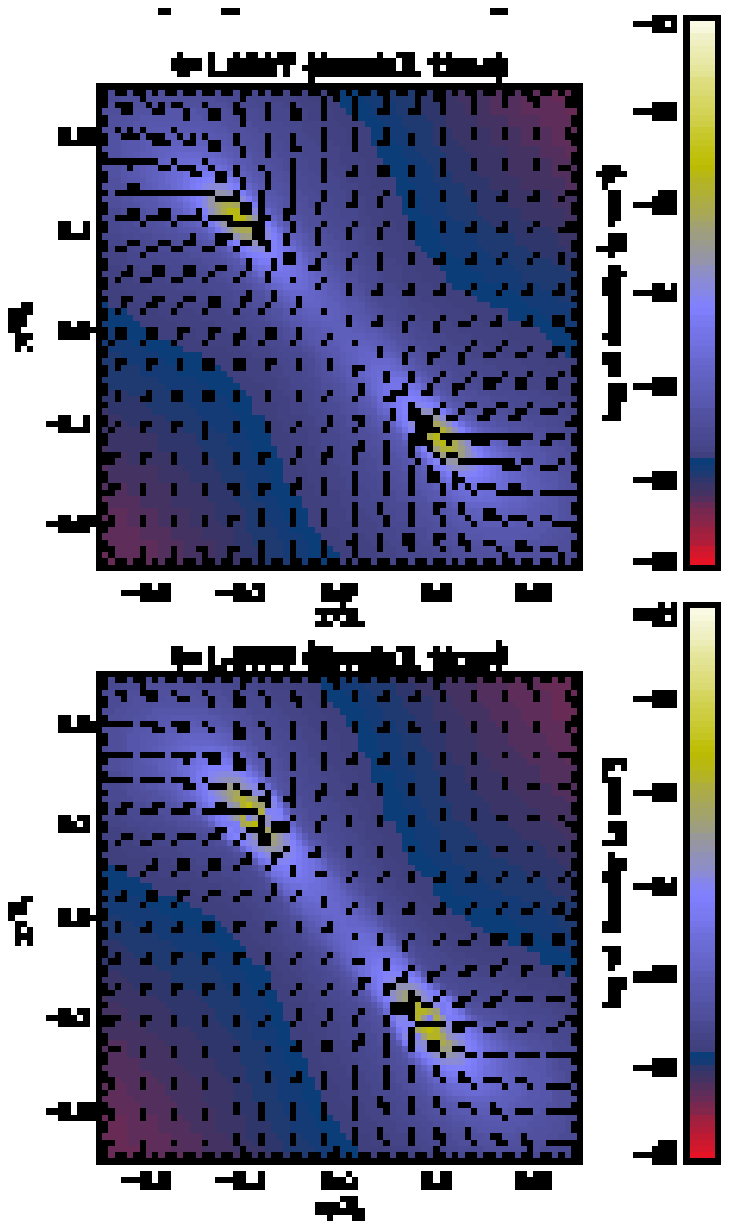}}
\end{picture}
\caption{Same as Fig.~\ref{mu1000_Ap0.1} for $\mu=2$ and $A=0.5$.}
\label{mu2_Ap0.5}
\end{figure}

\setlength{\unitlength}{1cm}
\begin{figure}
\begin{picture}(0,12)
%\put(0,6){\includegraphics[width=7cm]{mu1.25_0.5/dens_vit_xy00040.ps}}
%\put(0,0){\includegraphics[width=7cm]{mu1.25_0.5/dens_vit_xy00060.ps}}
\put(0,0){\includegraphics[width=7cm]{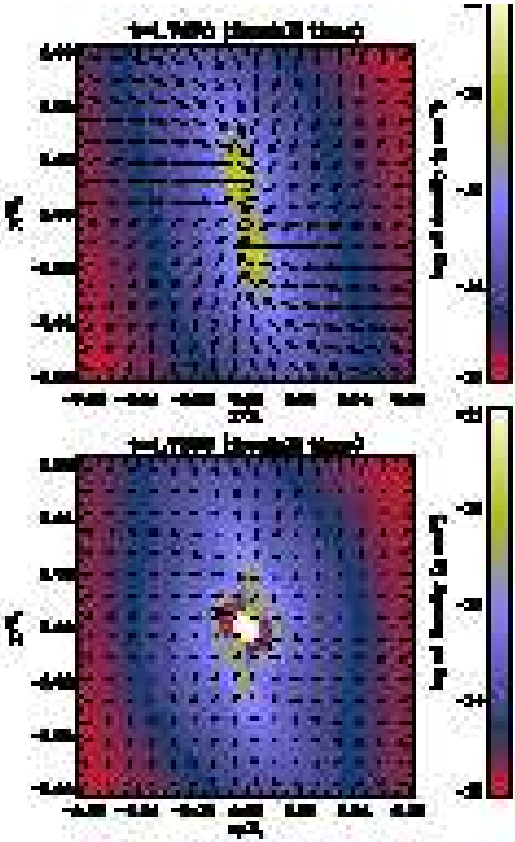}}
\end{picture}
\caption{Same as Fig.~\ref{mu1000_Ap0.1} for $\mu=1.25$ and $A=0.5$.}
\label{mu1.25_Ap0.5}
\end{figure}

\section{Strong initial perturbations}
In this section, we investigate the effect of a density fluctuation
having a larger initial amplitude, $A=0.5$. In that case the cloud is 
more prone to fragment because  the perturbations develop quickly.
Such strong perturbations can be due either to initial conditions 
or to strong external triggering. 
Although we are keeping in this section the same amount of rotation as
in the previous section, we also performed simulations with the same initial 
conditions but with no rotation and found similar results during the early phase
of fragmentation as in the case  with rotation. 
This indicates that with such strong perturbations, the initial cloud
fragmentation is purely thermal and independent of the  rotation. We  refer 
to this situation as thermal fragmentation. 

\subsection{Thermal fragmentation}
We present results for weak and intermediate values of magnetic strengths, 
namely $\mu = 20$ and $\mu = 2$.
Figure~\ref{mu20_Ap0.5} shows results for $\mu=20$. The first panel shows that 
a filamentary structure develops in which two protostar have formed. The two 
protostars are well separated by roughly $\simeq 0.15 \times R_0$ as opposed to 
$\simeq 0.01 \times R_0$ in the case of weaker perturbations. 
As explained previously, the fragmentation in this case has a purely thermal
origin  and is simply 
due to the development of the initial perturbations 
(in the case of weak perturbations, rotationally driven 
fragmentation does not strongly depend on the perturbation amplitude).
The second panel shows that the two fragments  are approaching each other falling along 
the filament.  The final panel shows that the two fragments have merged and that 
a protostellar disk forms. At this point, however, it is necessary to remember 
that the second collapse is not treated in the simulation and that only the 
first Larson core is considered. Therefore the merging should be considered with 
great care. Careful thermodynamical treatment of the first Larson core is required.
 Indeed the 
first core is known to have a short life time of roughly $10^3$ years. Therefore since 
the timeshift between the first and second panel is larger than $10^4$ years, it is 
expected that second collapse would have already occurred by the time, the two 
fragments approach each other. Since stars have very small radii, it is unlikely that 
the binary is going to merge.
Therefore the final product could be a binary rather 
than a single protostar. 

Figure~\ref{mu2_Ap0.5} shows results for $\mu=2$. Both panels show that 
two fragments develop in a similar  way as for $\mu=20$. Unfortunately, 
the relative strong value of the magnetic intensity, makes the time steps much 
shorter than in the previous case and it becomes computationally expensive to follow the long range evolution.
It seems however likely that a similar evolution is expected leading to the merging and 
the two first cores.

\subsection{Strong field: suppression of thermal fragmentation}
Figure~\ref{mu1.25_Ap0.5} shows results for $\mu=1.25$ and $A=0.5$. 
As revealed by the first panel, the $m=2$ density perturbation has 
only weakly developed and only two shallow density maximum are visible. 
The second panel shows  a single object  in the centre surrounded by 
a pseudo-disk. Since the timeshift between the two panels is about
 $0.015 $ freefall time ($\simeq 5 \times 10^2$ years) there is not sufficient time
for the second collapse to arise. It seems plausible that in this 
particular case, a single star is likely to form.

Therefore, it appears, that a nearly critically
 magnetized dense core, ideally coupled to the magnetic field, is 
difficult to fragment into a few objects even if relatively large fluctuations
are present initially.   This conclusion is in 
good agreement with the conclusion reached by Hosking \& Whitworth (2004), 
that a magnetically subcritical cloud, which collapses after
sufficient magnetic flux has been lost because of  ambipolar diffusion, 
does not fragment. \\

We conclude that, with large amplitude density fluctuations, a collapsing dense 
core fragments over a large range of initial conditions as long as it is not 
too strongly magnetized. The fragments are due to each initial seed collapsing 
individually. The question of the survival of the fragments should  however be 
carefully investigated. We note that Price \& Bate (2007) 
reach very similar conclusions. They also propose that in some cases, 
as when the magnetic field is perpendicular to the initial rotation axis, the magnetic field
may help the fragment to survive.

\section{Discussion}
Since there is no question that a large fraction of stars are binaries, the 
efficient stabilization  of the disk and the suppression of fragmentation even for
modest field amplitudes may constitute a severe difficulty. We have no  answer
to this apparent conundrum but we discuss three  possible mechanisms which may induce the fragmentation 
of the dense cores and the formation of binaries.  We also suggest various observational
tests which could discriminate between these scenarios.

\subsection{Large initial density fluctuations}
We start with the possibility of having sufficiently large fluctuations 
initially. As demonstrated in this paper and in Price \& Bate (2007), fragmentation
is possible in such a situation. The question is then how likely are  large 
initial perturbations ? 
Observationally the question is not easy to address. 
It is nevertheless known that the dense cores 
usually are not very uniform and that typically sonic velocity dispersion is observed. 
This is broadly compatible with the presence of  density perturbations although maybe not 
as large as 50$\%$. On the other hand, cores with a high aspect ratio are observed. If  some 
of them are elongated, i.e. prolate objects, this may be equivalent to a substantial $m=2$
perturbation. 
Indeed, simulations of  quiescent cores
having   transonic turbulence, present  significant 
fluctuations (Goodwin et al. 2004). It seems however difficult at this stage to infer 
quantitatively whether the perturbations generated by this weak turbulence are sufficient 
to produce multiple systems.

Another possibility is that the collapse may be induced by an external agent such as 
supernova remnant or protostellar jet. These are likely to generate
large perturbations. Indeed, all simulations considering clouds evolving far from equilibrium
(Bate et al. 2003, Ballesteros-Paredes et al. 2003, Hennebelle et al. 2006, Peretto et al. 2007)
do find dense cores with initially strong perturbations.

Future works should specify under which conditions
such initial fluctuations are likely to be produced. This may depend on the physical properties 
of the molecular cloud in which the dense core is embedded. For example, the answer could be different
in the Taurus molecular cloud in which star formation is relatively quiescent and in the Orion 
molecular cloud where star formation is more active. 

We stress that this mechanism constitutes a change of paradigm with respect to
 the standard hydrodynamical scenario (corresponding to Fig.~1).
  Indeed  this mechanism implies that the fragmentation is the result
of perturbations seeded at large scales rather than an intrinsic properties of the collapsing dense core. 
 
An important prediction of this  model is that 
the binary should not be necessarily located in the equatorial plane of the core since the initial
perturbations should likely be randomly distributed.

\subsection{Ambipolar diffusion}
In the present calculations,  perfect coupling between gas and magnetic field is assumed. 
The loss of flux due to  ambipolar diffusion could therefore possibly help to reduce
the magnetic intensity. 
The ratio of the ambipolar diffusion time, $\tau_{\rm ad}$, and the freefall time, $\tau_{ff}$, 
has been computed by Shu et al. (1987). When the magnetic field is just sufficient to compensate  gravity, 
they obtain $\tau_{\rm ad} / \tau_{ff} \simeq 8$, indicating that the ambipolar diffusion time is 
larger than the freefall time. When the magnetic field is weaker, the ambipolar time increases
further. Therefore, it appears unlikely that magnetic field could be removed from the envelope of 
a supercritical collapsing core.

However, the possibility remains that  ambipolar diffusion could occur in the disk allowing 
the toroidal magnetic field to be diffused out. To investigate this issue, we estimate 
the ratio of ambipolar diffusion time in the disk along the vertical direction and the 
growth time of the toroidal magnetic field given by Eq.~(\ref{growingtime}).
Here we write it as
\begin{eqnarray}
\tau_{{\rm mag}} \simeq { \sqrt{4 \pi \rho} h C_s \over    B_z  V_\theta    }. 
\label{growingtime2}
\end{eqnarray}

The ambipolar diffusion time of the toroidal component along the 
vertical direction is:
\begin{eqnarray} 
\tau_{\rm ad} \simeq { \gamma \rho \rho_i h ^2 \over B_\theta ^2}, 
\label{ambipol}
\end{eqnarray}
where $\gamma= 3.5 \times 10^{13}$ cm$^3$ g$^{-1}$ s$^{-1}$ is the drag coefficient
 and $\rho_i$ the density of ions.
Since we are interested in the value of $B_\theta$ such that 
$B_\theta / \sqrt{4 \pi  \rho} \simeq C_s$, the ambipolar diffusion time becomes:
\begin{eqnarray} 
\tau_{\rm ad} \simeq {\gamma  \rho_i h ^2 \over 4 \pi C_s ^2}, 
\label{ambipol2}
\end{eqnarray}

Following Shu et al. (1987), we write  $\rho_i = C \sqrt{\rho}$
where $C=3 \times 10^{-16}$ cm$^{-3/2}$ g$^{1/2}$.
With $h \simeq r / \sqrt{2 d}$, $V_\theta = \sqrt{M_s G /r}$
and $B_z= H_z C_s^2 / \sqrt{G} r$, we obtain
\begin{eqnarray}
{ \tau_{\rm ad} \over \tau_{\rm mag} }  = { \gamma \; C \over (4 \pi)^{3/2}  C_s } \sqrt{{ M_s \over r }}  {H_z \over \sqrt{2d}}.   
\label{ratio}
\end{eqnarray}
Thus, we obtain
\begin{eqnarray}
{ \tau_{\rm ad} \over \tau_{\rm mag} }  = 
3 \times \sqrt{ M_s \over 0.1 M_\odot} \sqrt{ 100 {\rm AU} \over  r }  {H_z \over \sqrt{d}}.   
\label{ratio2}
\end{eqnarray}
For  $\mu=20$, $H_z \simeq  0.4$. From  Fig.~2 of  paper I,  $d \simeq
10$. Thus, the  ambipolar diffusion time is comparable  to the growth
time of the toroidal component.  It is therefore likely  ambipolar
diffusion will change  the picture  slightly. It may  help clouds  with low
$\mu$ to fragment more easily. We do not expect it however, to shift very significantly
the value of $\mu$ for which fragmentation is occurring. Since $H_z$ is proportional
to $1/\mu$, ${ \tau_{\rm ad} / \tau_{\rm mag} }$ increases rapidly when $\mu$
decreases.

\subsection{Fragmentation during the second collapse}
Fragmentation during the second collapse has been investigated by Bonnell \& Bate (1994)
with the prospect of exploring whether the formation of close binary systems could be possible. 
They conclude that the fragmentation is indeed possible but that the binaries have to accrete most 
of their mass since they are initially very small.  

On the other hand, Machida et al. (2007a) recently explored the magnetized second collapse
taking into account the large Ohmic dissipation which is predicted to occur in the first Larson core by various models 
(e.g. Nakano et al. 2002). As a result, most of the magnetic flux is lost making  the fragmentation easier
to proceed. Indeed, the very recent study of Machida et al. (2007b) shows that fragmentation during 
second collapse is not only possible but a promising mechanism.
Interestingly, Banerjee \& Pudritz (2006) also reports the formation of a very close binary in 
their second collapse calculations despite the ideal mhd assumption.  
It therefore appears possible that fragmentation could occur during this phase possibly 
driven by rotation.  In this case, the binary should gain sufficient angular momentum from the accretion 
to increase the separation between the two stars (see e.g. Goodwin et al. 2004).

The prediction of this model is that the binaries should be in the equatorial plane since the 
orbital angular momentum of the binary is due to the angular momentum of the accreting gas.
Also the separation between the 2 stars should increase with time implying that closer binaries
should be observed on average into younger cores.
Note that this mechanism could possibly 
work even if the magnetic field is initially very strong in the core.

In order to test each of these scenarios, high resolution observations
would be necessary. In this respect, ALMA will certainly be a very powerful tool
as demonstrated by the synthetic observations performed by Andr\'e et al. (2007) using the 
present calculations.

\section{Conclusion}
In this paper  we have studied the fragmentation of a collapsing 
magnetized molecular dense core by performing a set of numerical 
simulations. We interpret the results with the analysis performed 
in the companion paper in which we investigate the accretion and ejection 
processes that take place in the simulations. 

With our choice of initial conditions, 
 and for  perturbations of  low amplitude, unmagnetized cores
fragment, while magnetized  cores having values of  $\mu$ as large as 20  
 do not fragment. Based on an analytic estimate, 
we suggest that for cores having $\mu$ larger 
than $\simeq 15 / \alpha$,  the growth rate of the toroidal component is 
too slow to stabilize the disk.  
We stress that the suppression of fragmentation in this range of 
parameters, is not due to magnetic
braking  but to  the rapid  growth of  the toroidal  component  of the
magnetic  field  induced  by  the  differential  rotation  within  the
disk. The Alfv\'en speed associated to this toroidal component adds up
to the sound speed of the disk and stabilize it.

For  values of $\mu$ smaller than $5$, no big centrifugal disk forms because
first the collapse occurs mainly along the field lines bringing less angular momentum 
and second  magnetic braking  removes  angular momentum. This makes
these cores even less prone to fragment. 

The situation is different if large amplitude perturbations are initially seeded. In this case, 
each perturbation develops independently even without rotation. Since the strong field 
amplification is primarily due to differential motions in the disk, the 
magnetic field is unable to suppress fragmentation except if the core is almost 
critical, i.e. the field is initially very strong. 
The following evolution of these fragments requires a careful treatment of the thermodynamics
of the first Larson core. It is indeed likely that if the second collapse has not occurred 
by the time where these fragments approach each other, they are going to merge. On the other hand, if the 
protostars already formed, the binary system is likely to survive.

For dense cores having rotation and magnetic strength typical of values inferred from observations, 
we find that fragmentation is suppressed by the magnetic field if the initial density perturbations
are too small. This constitutes a severe problem, since there is no question that a significant 
fraction of stars are binaries. 
In view of this, we discuss the likelihood of having sufficient perturbations within the cores
initially, the impact of ambipolar diffusion and the possibility of fragmenting during
the second collapse phase. We speculate that the first may depend on the physical characteristic of the
 molecular cloud in which the dense core is embedded  whereas the last should be relatively independent 
of the large scales and could work even if the magnetic field is initially very strong.

\acknowledgement
Some of the simulations
presented in this paper were performed at the IDRIS supercomputing center and on the CEMAG 
computing facility supported by the French ministry of research and education
through a  Chaire d'Excellence awarded to Steven Balbus.
We thank S\'ebastien Fromang, Doug Johnstone and Philippe Andr\'e for a critical reading of the manuscript. 
We thank Frank Shu, the referee, for helpful comments. 
PH thanks Masahiro Machida for related discussions.

\end{document}